Predicting synthesizable manganese nitride with unprecedentedly giant magnetocrystalline anisotropy energy

Ze-Jin Yang[1],[*]

[1]School of Science, Zhejiang University of Technology, Hangzhou, 310023, China

Using modern crystal structure prediction program (CALYPSO), we searched many experimentally synthesizable low-energy structures with perfect or nearly perfect easy-axis magnetocrystalline anisotropy energy (MAE) in manganese nitride, including MnN, $Mn_2N$, $Mn_3N_2$, $Mn_5N_2$, $Mn_4N$, respectively, which are the more frequently studied stoichiometries by experimental researchers. MnN ( $I\bar{4}2d$ ) shows giant MAE with values of $E_{001}$=1006, $E_{010}$=0, $E_{100}$=920 $\mu$eV/atom (same hereinafter), respectively. One perfect easy-axis MAE in $Mn_3N_2$ ($P4_2/mmc$) with correspondent values of $E_{010}$=$E_{100}$=12 is observed, the other nearly perfect easy-axis MAE one (*Ibam*) with respective values of $E_{001}$=324 and $E_{010}$=345 is observed. Four almost totally perfect easy-axis MAE structures are obtained in $Mn_2N$, including *P4/mmm* with individual $E_{001}$=249 and $E_{100}$=250, *Pccm* with $E_{001}$=$E_{100}$=62, *P4/nmm* with $E_{001}$=58 and $E_{100}$=60, *Imma* with $E_{001}$=108 and $E_{100}$=109, respectively. Three structures including one perfect candidates are found in $Mn_4N$, including *Fmmm* with individual $E_{001}$=126 and $E_{010}$=121, *I4/mmm* with $E_{010}$=127 and $E_{100}$=133, *I4/mmm* with $E_{001}$=$E_{100}$=169, respectively. Too many valuable structures are deserved to be further studied by both theoretical and experimental scientists. The present study might attract close attention to these several compounds.

---

[*] zejinyang@zjut.edu.cn

## I. Introduction

Permanent magnets with extraordinary magnetic properties are extensively used in modern industry such as in computer hard disk drives and wind turbine generators as well as electric or hybrid car motors, these properties mainly includes large Curie temperature ($T_c$) and saturation magnetization ($\mu_0 M_s$) as well as the magnetocrystalline anisotropy energy (MAE), particularly for those uniaxial ($K_u$) or easy-magnetization-axis MAE compounds. MAE defines the energies or magnetic work required for bringing the direction of the magnetization from the easy direction to that imposed by the performed magnetic field (vice versa, it is the energy that forces the moments to align along the easy axis, or the energy required to reorient the electron spins in a ferromagnet from easy- to hard-axis directions). $SmCo_5$ and $NdFe_{14}B$ are the two most prominent substances in commercial production, while both of them contain rare-earth elements which are very expensive in price due to their very limited resource in our earth. For the more extensive modern industry applications, the researchers have to search or synthesize more cheap materials with same properties.

Mn-N compounds exist rich magnetic phases, such as until now, the ε-$Mn_4N$, ζ-$Mn_2N$ (frequently called $Mn_5N_2$), θ-MnN, and η-$Mn_3N_2$ phases have been synthesized[1-5], also including θ-$Mn_6N_{5.26}$ *etc*[6], whereas their MAE is less studied due probably to the practical difficulties whether in theoretical computation or experimental measure, thus it is important to explore whether there exists the excellent MAE in Mn-N compounds. Moreover, considering the possible difficulty to synthesize the above alloys compounds, it is also very difficult to synthesize the perfect stoichiometry, it is necessary to theoretical prediction them to guide experimental exploration in order to reduce the cost. Mn is very cheap relative to that of Co, study on Mn is thus absolutely necessary. Based on these motivations, we carefully searched and analysized the obtained structures for MnN, $Mn_2N$, $Mn_3N_2$, $Mn_5N_2$, $Mn_4N$, respectively, many papers have discussed the structures and the properties of the above alloys, we thus ignore to repeat them for simplicity. For easier reading purpose, we also ignore the decimal data for MAE at the discussion section.

## II. Methods

The structures are searched by CALYPSO code[7,8], and further optimized by projector-augmented wave method[9]. The spin polarized generalized gradient approximation and the Perdew-Burke-Ernzerh of functional are used to model the exchange-correlation energy[10]. The electronic and force convergence is below $10^{-7}$ eV and 0.01 eV/Å, the k-point density is below $2\pi \times 0.02$ Å$^{-1}$, and plan-wave basis is 350 eV, which are sufficient for the magnetic atoms based on the calculations on iron silicides[11]. The on-site coulomb interaction term $U$ (and $J$) is ignored based on the previous conclusion of iron silicides[11].

## III. Results and discussions

**MnN**

All of the lowest-energy axes of the structures are set to the energy reference and ignored for simplicity. The lowest-energy MnN structure (ID_1,2,3, Table 1) has a symmetry of $F\bar{4}3m$ (No.216), consisting with previous calculations [12].

Furthermore, *I*4/*mmm* (No.139) or $Fm\bar{3}m$ (No.225) is the metastable structure, also agreeing with previous synthesization[12]. The $F\bar{4}3m$ structure presents $E_{010}$=454, $E_{100}$=624 $\mu$eV/atom, respectively, which is very large MAE despite its non-perfect easy-axis distributions. Several hundreds of $\mu$eV/atom is a giant value that is not usually occurred in the materials. Owing to the practical application only concerns the total MAE value, we thus ignore to discuss the spin and orbital moment components. Such large values still have certain applications in some special technique fields requiring multi-axis step-like energy distribution features. This $F\bar{4}3m$ might be easily synthesized due to its perfect cubic structure, in which each Mn has four first-nearest neighbors N, and each N also has four first-nearest neighbors Mn, similar to that of methane. In addition, the lattice energy will be lowered (from $E_f$=-0.2523 to -0.2639 eV/atom) if the lattice parameter elongates slightly (from 4.2074 to 4.2376 Å), as a result the axial energy order and value also redistribute, with values of $E_{001}$=642, $E_{010}$=347 $\mu$eV/atom, respectively. This phenomenon illustrates the strong atomic distance dependence of the MAE under the lattice parameter elongation of less than about 0.7%. Negligible lattice parameter elongation in ID_3 relative to that of ID_1 will alternate axial energy order, revealing the strong atomic site dependence of spin-orbital coupling.

A slightly distorted structure (ID_4, *Cm*, No. 8, $E_f$=-0.2582 eV/atom), whose MAE is reduced correspondingly to $E_{001}$=262, $E_{010}$=200 $\mu$eV/atom. Its atomic arrangement is generally same with those of ID_1-3, just as whose symmetry is $F\bar{4}3m$ when determining it by a tolerance of 0.1 Å. To obtain the site dependence of the MAE, the more usually used method is to shift the atom gradually while keeping the other atoms unchanged, which will cause the lattice instable and induce substantial stress within the lattice more or less, corresponding to a hypothetical and transient state structure, and such strained structure is impossible to be synthesized experimentally. However, the present similar structures show multiple and simultaneous variations in the whole lattice parameters and correspond to a stable state, comprehensively illustrated the complexity of the spin-orbital coupling. All of the studied similar structures might be synthesized with non-equilibrium technique. Therefore, the present study is more practicable and useful.

The experimental[4] NaCl-fct (face-centered tetragonal) MnN has almost fcc (face center cubic) structure and relatively higher energy (ID_5, $E_f$=-0.1737 eV/atom) than that of $F\bar{4}3m$ (No.216, $E_f$≈-0.25 eV/atom), whose MAE are $E_{001}$=48, $E_{100}$=87 $\mu$eV/atom, respectively. However, a perfect fcc structure will produce giant MAE in ID_6, it is $E_{001}$=2388, $E_{100}$=124 $\mu$eV/atom, respectively.

The searched lowest-energy structure (ID_7), with a symmetry of $F$222 (No. 22) and $E_f$=-0.265 eV/atom, has zero magnetic moment and MAE. The other several structures, with energy ranges of about -0.21~-0.25 eV/atom, have generally same atomic distributions and large MAE with the only one exception (ID_8, $E_f$=-0.2399 eV/atom, *Pmmm*, No.59) that shows the negligible MAE, such negligible MAE is due probably to its special atomic first-nearest neighbor coordinate number (CN), namely, the CN of Mn and N are 6. In other words, all of the first nearest neighbors of Mn are N, vice versa, which reduced the spin-orbital coupling strength and thus canceled the magnetism. Many structures have the methane-like building block, in which the Mn and N have the same four CN value, such atomic arrangement will leave sufficient space for the neighboring Mn interaction. In other words, the average angle between the two neighbor Mn-N is far larger than that of CN=6 (about 90°). Such as the average values of the four angles of the N-Mn-N within $MnN_4$ building block is about 110°.

A structure (ID_9), with $E_f$=-0.212 eV/atom and symmetry $I\bar{4}2d$ (No.122), about 40 meV/atom higher than that of lowest-energy structure, shows unprecedentedly giant nearly perfect easy-axis MAE, with values of $E_{001}$=1006, $E_{100}$=920 $\mu$eV/atom, respectively, which is very possible to be synthesized experimentally and has very promising applications. This giant value indicates that atomic arrangement plays a key role during the spin-orbital coupling despite that the chemical stoichiometry between magnetic and non-magnetic atoms is 1:1. In addition, a slight distortion structure (ID_10) including an elongation of 0.1 Å along $c$ axis and shrinkage 0.005 Å along $a$ and $b$ axes will decrease 16 meV/atom in energy and produce $E_{010}$=1030, $E_{100}$=131 $\mu$eV/atom, respectively, such small variation doesn't change the structure symmetry, which further demonstrates that the MAE is quite sensitive on the atomic site. In a word, it is clear that the lattice distortion unavoidably enhanced the lattice energy, whereas it also enhanced the MAE in many cases. In fact, many metastable phases usually present large MAE than that of ground state. Meanwhile, the available ID_11 presents similar MAE with that of ID_10. Conclusively, MnN is an important compound that consists of many low-energy synthesizable structures with large nearly perfect easy-axis MAE. Usually, magnetic transition temperature is linearly dependent on the magnetic moment magnitude[13] in binary manganese nitrides, large moment usually induces large MAE, whereas the present MnN does not obey the general empirical rule.

Table 1 Experimental and searched low-energy MnN structures information, formation of energy ($E_\text{f}$, eV/atom), magnetocrystalline anisotropy energy (MAE, $\mu$eV/atom), identification number (ID), symmetry determination under different tolerances (Å), lattice parameters and angles, $a$, $b$, $c$, (Å), $\alpha$, $\beta$, $\gamma$, (°), the words "Experimental" and "Searched" have been abbreviated as "Exp." and "Sea" for simplicity, respectively. The 'available' means that the coordinate data is downloaded from www.materialproject.org. The big (purple) and small (blue) balls are Mn and N atoms, respectively, the same is applicable in the following tables.

| ID<br>$E_\text{f}$<br>Exp./Sea. | MAE<br>001<br>010<br>100 | Sym.<br>0.001<br>0.01<br>0.1 | Unit cell<br>●Mn<br>●N | $a,b,c,$<br>$\alpha,\beta,\gamma,$ |
|---|---|---|---|---|
| 1<br>-0.2523<br>available | 0<br>454.006<br>624.335 | $F\bar{4}3m$ 216<br><br>$F\bar{4}3m$ 216<br><br>$F\bar{4}3m$ 216 | 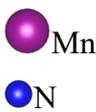 | 4.20745<br>4.20745<br>4.20745<br>90,90,90 |
| 2<br>-0.2639<br>Sea. | 642.9137<br>347.02750 | $F\bar{4}3m$ 216<br><br>$F\bar{4}3m$ 216<br><br>$F\bar{4}3m$ 216 | 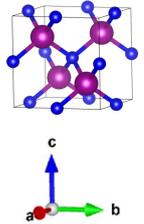 | 4.23763<br>4.23763<br>4.23763<br>90,90,90 |
| 3<br>-0.252<br>Sea. | 579.391<br>206.887<br>0 | $F\bar{4}3m$ 216<br><br>$F\bar{4}3m$ 216<br><br>$F\bar{4}3m$ 216 | 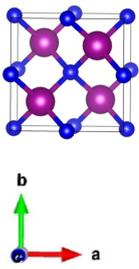 | 4.20785<br>4.20785<br>4.20785<br>90,90,90 |
| 4<br>-0.2582<br>Sea. | 262.877<br>200.112<br>0 | $Cm$,8<br>$Cm$,8<br>$F\bar{4}3m$ 216 | 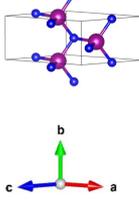 | 5.23877<br>2.95821<br>4.19565<br>90<br>145.0479<br>90 |
| 5<br>-0.1737<br>NaCl-fct<br>Exp. | 48.0325<br>0<br>87.335 | $I4/mmm$ 139<br><br>$Fm\bar{3}m$ 225<br><br>$Fm\bar{3}m$ 225 | 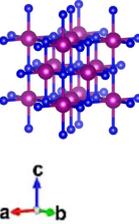 | 4.05778<br>4.05778<br>4.05002<br>90,90,90 |
| 6<br>-0.1976<br>available | 2388.3<br>0<br>124.68 | $Fm\bar{3}m$ 225<br><br>$Fm\bar{3}m$ 225<br><br>$Fm\bar{3}m$ 225 | 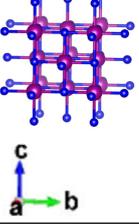 | 4.06495<br>4.06495<br>4.06495<br>90,90,90 |
| 7<br>-0.265<br>Sea. | 0<br>0<br>0 | F222, 22<br>$I\bar{4}m2$ ,119<br>$I\bar{4}m2$ ,119 | 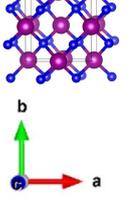 | 4.12575<br>4.38498<br>4.12459<br>90,90,90 |
| 8<br>-0.2399<br>Sea. | 7.85<br>0<br>2.265 | $Pmmm$, 59<br>$Pmmm$, 59<br>$Pmmm$, 59 | 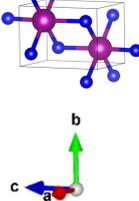 | 2.89174<br>2.80830<br>3.94227<br>90,90,90 |
| 9<br>-0.2123<br>Sea. | 1006.19<br>0<br>920.2125 | $I\bar{4}2d$ ,122<br><br>$I\bar{4}2d$ ,122<br><br>$I\bar{4}2d$ ,122 | 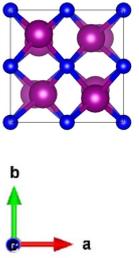 | 4.32858<br>4.32858<br>7.86636<br>90,90,90 |

| 10 -0.2382 Sea. | 0 1030.9 131.762 | $I\bar{4}2d$, 122 <br> $I\bar{4}2d$, 122 <br> $I\bar{4}2d$, 122 | 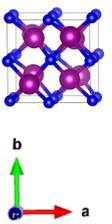 | 4.32373 4.32373 7.95674 90,90,90 |
|---|---|---|---|---|
| 11 -0.1987 available | 185.495 1313.07 0 | $P6_3mc$ 186 <br> $P6_3mc$ 186 <br> $P6_3mc$ 186 | 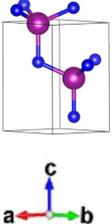 | 2.91368 2.91368 5.21045 90,90,120 |

**Mn$_3$N$_2$**

Many searched Mn$_3$N$_2$ structures have comparable energies within the $E_f$ range of -0.2~-0.21 eV/atom, in which the lowest-energy structure of Mn$_3$N$_2$ has a $E_f$ of about -0.21 eV/atom. In fact, almost all of them have nearly identical structures or lattice shapes ($\alpha=\beta=\gamma=90°$) except for some minor difference in lattice parameters or atomic coordinates, which might be one of the reasonable origins of their comparable MAE along the *a* (100), *b* (010), *c* (001) axes, respectively. In detail, the MAE are about $E_{001}\approx110$, $E_{100}\approx80$ $\mu$eV/atom, respectively. The common symmetry might be *I*4/*mmm* (No.139) when a tolerance or criterion of 0.01 Å is used, whereas *P*4$_2$/*mcm* (No.132) or *P*4$_2$/*mnm* (No.136) might be obtained under 0.001 Å originating from the subtle lattice distortion, as is clearly seen from the variations of the lattice parameters *a* and *b* with a fluctuation of ≥0.01 Å, see structures ID_1 and ID_2 (shown in Table 2) and ID_S1-3 shown in Table S1 of supplemental information (SI) for details. Therefore, it is possible to synthesize this structure as there always exists a state at this energy range, in addition to its medium magnetism is also useful in some special fields.

A lowest-energy Mn$_3$N$_2$ structure (ID_2, Table 2) is searched and shows decreased easy-plane MAE, $E_{001}$≈61, $E_{100}$≈0.6 $\mu$eV/atom, respectively. Its symmetry is *Cmcm* (No. 63) at a tolerance of 0.01 Å, or *I4/mmm* (No.139) at a tolerance of 0.001 Å, respectively. The spin-orbital coupling strength is totally reduced at all the three axes when comparing its correspondent axial energy to those of ID_1. Therefore, this is a convincing example to justify the fact that the lattice distortion indeed could substantially enhance or reduce the MAE even under the same lattice shape with only certain lengthening and shortening along the lattice axis.

A Mn$_3$N$_2$ structure (ID_3) shows $E_f$≈-0.2 eV/atom, $E_{001}$≈92, $E_{100}$≈192 $\mu$eV/atom, respectively, whose symmetry is *Pnma* (No. 62) at a tolerance of 0.01 Å, or *Pbam* (No.55) at a tolerance of 0.001 Å, respectively. Clearly, this structure also has relatively large distortion with those of ID_1 and ID_2, whose energy is about 10 meV/atom higher than that of ID_2. The currently searched diverse MAE could enrich our understanding to the coupling mechanism and guide the experimental synthesization.

The $E_f$ of the searched Mn$_3$N$_2$ structure (ID_4, *Cmcm*, No. 63) is about -0.2 eV/atom, whereas its magnetic moment is zero, presenting an isotropy energy distribution and the cancellation of the spin and orbital moment, clearly, this kind of structure has direct connection with those of ID_1 and ID_2 according to the hexagonal structure feature.

The other energetically comparable Mn$_3$N$_2$ structures show a $E_f$ range of about -0.17~-0.18 eV/atom, about 30 meV/atom higher than those of -0.2~-0.21 eV/atom discussed above. Structures with energy differences below about several meV usually belong to the same phase originating from the very similar atomic arrangement. Thus, these structures provide valuable references for comparison the structure-MAE relationship as candidate populations locating at different potential energy valleys, similarly, it is also a good idea to catalogue and discuss them together.

One of the stoichiometry (the coordinate files could be downloaded from www.materialproject.org) $Mn_3N_2$ phase (ID_5) has a $E_f$ of -0.156 (*I4mmm*, No.139), which presents large MAE, including $E_{001}$≈360, $E_{010}$≈206 $\mu$eV/atom, respectively. ID_6 shows a contraction along the three axes compared with that of ID_5, whereas its MAE remains almost unchanged, indicating an isotropic contraction. Meanwhile, a small lattice shortening in ID_6 didn't change the energy order and values among the three axes relative to that of ID_5, differing with the case of ID_1 of MnN in Table 1. ID_5 has a similar unit cell with those of structures having $E_f$≈-0.2~-0.21 eV/atom, such as ID_1 and ID_2. As far as the large MAE values are concerned, ID_5 might also has some potential application despite its non-perfect easy-axis MAE nature. The other experimental one (ID_7, $P\bar{3}m1$, No.164) is a high-energy state, whose $E_f$≈-0.125 eV/atom and MAE is small, in detail, $E_{001}$≈21, $E_{010}$≈10 $\mu$eV/atom, respectively. The other similar structures with that of ID_5 is shown in Table S1-2.

Structural tension or compression indeed could adjust MAE, such as one $Mn_3N_2$ structure (ID_8) shows $E_f$=-0.17 eV/atom, with a symmetry of $P6_3/mcm$ (No.193), whose MAE is about 7, 10, 0 $\mu$eV/atom along 001 (*c*) and 010 (*b*) direction relative to that of 100 (*a*) direction, respectively. Slight elongation along *a*, *b*, *c* directions ($a=b$=4.6214(4.6432), $c$=9.1971(9.2236) Å) with a value of about 0.02 Å will enlarge the corresponding MAE to about 29, 74, 0 μeV/atom with the cost of enhancement of lattice energy about 2 meV/atom, as is shown in Table S3 (ID_S4). Similarly, the MAE of ID_S5 (Table S4) will be about 98, 0, 87 μeV/atom but the lattice energy remains almost unchanged when the lattice parameters change to $a=b$=4.6393, $c$=9.1192 Å, $α=β$=90°, $γ$=120.0001°, respectively. Roughly speaking, these structures show generally similar atomic arrangements. Strictly speaking, the space group of the three structures are different, such as ID_8 has most symmetry operation and thus shows a group of $P6_3/mcm$ (No.193) at any three different kinds of tolerances including 0.001, 0.01, 0.1 Å, respectively. It still is $P6_3/mcm$ at a precision of 0.1 Å in ID_S4 (Table S3), whereas it is $P6_3cm$ (No.185) at tolerances of and 0.001 and 0.01 Å, meaning that the atomic relative shift range is larger than 0.1 Å and corresponds to about 5% elongation of the chemical bonds. This is true in the case of ID_S5 (Table S4), whose symmetry is $P6_3$ (No.173) at a tolerance of 0.001 Å or $P6_32_2$ (No.182) at tolerances of 0.1 and 0.01 Å, respectively. The negligible variation is deserved to be carefully analysized as the electric device often works at high temperature conditions, originating probably from the large current intensity or the long working time or the slow heat dissipation and so on, which will inevitably enlarge the lattice size and thus change the MAE. The other similar cases are shown in Tables S3-4 for reference purpose.

A nearly perfect easy-axis MAE structure (ID_9) is searched, as is shown in Table 2, whose $E_f$=-0.137 eV/atom, $E_{001}$≈325, $E_{010}$≈345 μeV/atom, respectively. Its energy is about 20 meV/atom (*Ibam*, No.72) higher than that of experimental *I4/mmm* (No.139, $E_f$=-0.156 eV/atom), thus it is still possible to be synthesized experimentally. Similarly, a large MAE structure (ID_10) is also searched with a symmetry of *P2$_1$* (No.4), whose $E_f$=-0.12 eV/atom, $E_{001}$≈381, $E_{010}$≈307 μeV/atom, respectively. A perfect easy-axis MAE structure (ID_11) is searched, whose $E_f$=-0.129 eV/atom, $E_{010}$=$E_{100}$=12 μeV/atom, respectively, corresponding to a symmetry of *P4$_2$/mmc* (No.131) at a tolerances of 0.1 and 0.01 Å, whereas it is *Cccm* (No.66) at a tolerance of 0.001 Å, respectively, presenting some applications in the low-MAE requirements fields. In a word, Mn$_3$N$_2$ deserved to be synthersized as there always exists useful MAE at different energy states.

Table 2 Experimental and searched low-energy Mn$_3$N$_2$ structures information,

| ID<br>$E_f$<br>Exp./Sea. | MAE<br>001<br>010<br>100 | Sym.<br>0.001<br>0.01<br>0.1 | Unit cell<br>●Mn<br>●N | $a,b,c,$<br>$α,β,γ,$ |
|---|---|---|---|---|
| 1<br>-0.2121<br>Sea. | 102.86<br>0<br>77.645 | *P4$_2$/mnm*,136<br>*I4/mmm*,139<br>*I4/mmm*,139 | | 3.98585<br>3.98585<br>11.61646<br>90,90,90 |
| 2<br>-0.2148<br>Sea. | 61.74<br>0<br>0.635 | *Cmcm*,63<br>*I4/mmm*,139<br>*I4/mmm*,139 | | 3.98358<br>3.98358<br>11.62871<br>90,90,90 |
| 3<br>-0.2033<br>Sea. | 92.5<br>0<br>192.445 | *Pnma*,62<br>*Pbam*,55<br>*Pbam*,55 | | 6.87030<br>5.71719<br>4.90334<br>90,90,90 |
| 4<br>-0.2019<br>Sea. | 0<br>0<br>0 | *Cmcm*,63<br>*Cmcm*,63<br>*Cmcm*,63 | 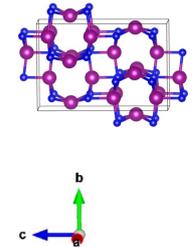 | 2.58051<br>6.69912<br>9.86924<br>90,90,90 |
| 5<br>-0.1566<br>Sea. | 360.789<br>206.156<br>0 | *I*4/*mmm*,139<br>*I*4/*mmm*,139<br>*I*4/*mmm*,139 | 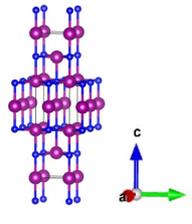 | 2.86260<br>2.86260<br>12.05071<br>90,90,90 |
| 6<br>-0.1572<br>Exp. | 361.671<br>206.488<br>0 | *I*4/*mmm*,139<br>*I*4/*mmm*,139<br>*I*4/*mmm*,139 | 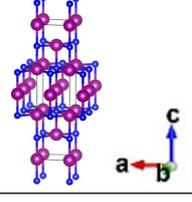 | 2.828973<br>2.828973<br>11.82875<br>90,90,90 |
| 7<br>-0.1254<br>Exp. | 21.442<br>10.378<br>0 | $P\bar{3}m1$,164<br>$P\bar{3}m1$,164<br>$P\bar{3}m1$,164 | 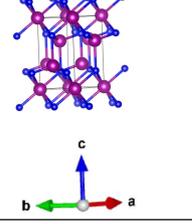 | 3.05375<br>3.05375<br>5.88195<br>90,90,120 |
| 8<br>-0.1701<br>Sea. | 7.095<br>10.82<br>0 | $P6_3/mcm$,193<br>$P6_3/mcm$,193<br>$P6_3/mcm$,193 | 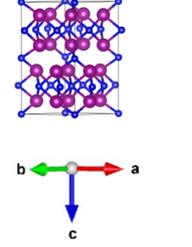 | 4.62140<br>4.62140<br>9.19712<br>90,90,120 |
| 9<br>-0.1369<br>Sea. | 324.075<br>345.485<br>0 | *Ibam*,72<br>*Ibam*,72<br>*Ibam*,72 | 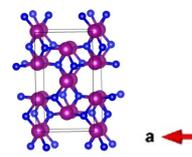 | 4.43632<br>5.12811<br>7.66036<br>90,90,90 |
| 10<br>-0.1202<br>Sea. | 381.455<br>307.855<br>0 | $P2_1$,4<br>$P2_1$,4<br>$Cmc2_1$,36 | 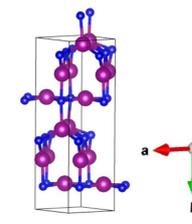 | 3.80226<br>11.81158<br>3.80017<br>90，92.3100，90 |
| 11<br>-0.1293<br>Sea. | 0<br>12.61<br>12.61 | *Cccm*,66<br>*P4$_2$/mmc*,131<br>*P4$_2$/mmc*,131 | 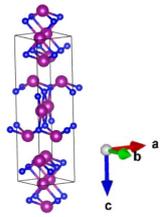 | 3.59118<br>3.59411<br>12.99978<br>90,90,90 |

**Mn$_2$N**

As is shown in Table 3, experimental Mn$_2$N *Pbcn* (No.60) has lowest energy among all of the studied structures, whereas different synthesization techniques usually obtain different structures due mainly to the complicated chemical reaction, as is seen from ID_S1-6 (Table S5). For clear purpose, we plot all of these candidate states in Figure 1. Unfortunately, these structures haven't very large magnetism and perfect easy-axis MAE. The axial energy order depends on the three lattice parameters among ID_S1-6 but the axial energy differences are almost unchanged, meaning that the orbital and spin moments are not significantly changed. We show these structures in order to emphasize that the low-energy Mn$_2$N presents not large MAE.

As is shown in Table 3, the $E_f$ and symmetry of ID_2 are -0.2451 eV/atom and *P6$_3$/mmc* (No.194), respectively, with MnN$_3$-type building block (tetrahedron). The correspondent coordinate number (CN) of Mn (or MnN$_3$) and N (or Mn$_6$N octahedron) is 3 and 6, the distance between Mn and N is about 1.95 Å, in addition, the distance between two nearest neighboring Mn is about 2.43 Å. This structure shows the layered-like structure with relative weak interaction between the two nearest neighboring Mn. The angles of the three N-Mn-N connection are 83.7°. Projections towards *a* and *b* axes are identical and thus their axial energies are also identical, Mn distributes at the center of an equilateral triangle formed by the three N atoms if the projections of Mn is towards *c* axis or *a*, *b* plane, thus the magnetism along *c* axis is also zero, the projection orientation of the nearest-neighbor single layer towards *a* or *b* axis is just contrary, thus the magnetism also canceled according to the structure feature of *P*6$_3$/*mmc*.

ID_3 ($E_f$=-0.2457 eV/atom, *P*2$_1$2$_1$2$_1$, No.19) shows medium MAE with values of $E_{001}$=36, $E_{100}$=119 μeV/atom, respectively. This low-symmetry structure causes the diverse distance between the Mn$_5$-N (within one unit cell), namely, 1.9404, 1.8421, 1.9907, 1.8686, 1.9279 Å, respectively.

Easy-axis MAE is the most ideal candidate material and could present extremely important industrial applications. The searched structure ID_4 (*Fmmm*, No.69) is a promising structure, as is shown in Table 3, whose $E_f$ is -0.203 eV/atom and MAE are $E_{001}$=249, $E_{100}$=250 μeV/atom, respectively, in addition, its MnN$_3$ building block is bend other than planar structure (the four atoms lay in a plane). A similar structure (Table 3, ID_5) with that of ID_4 shows $E_f$=-0.2295 eV/atom (*P*4$_2$/*mnm*, No.136) and medium MAE, including $E_{001}$=10, $E_{100}$=37 μeV/atom, respectively. Usually, electric device operates at high temperature and thus the lattice might be expanded inevitably or anisotropically, thus ID_4 might produce totally perfect easy-axis MAE at high temperature conditions.

Strictly speaking, many structures show $E_f$ within -0.19~-0.2 eV/atom with diverse symmetries, whereas they are essential the same structure due mainly to the more or less distortions. Roughly speaking, they have the symmetry $P4_2/mnm$ (No.136) with a tolerance of 0.01 Å. These structures show diverse but medium even small MAE, whereas no perfect easy-axis MAE is formed, as are shown in Tables S6-S8, ID_S7-19.

One structure (Table 3, ID_6) having $E_f$=-0.173 eV/atom and symmetry $Pccm$ (No. 49) shows almost perfect easy-axis MAE with medium values of $E_{001}$=62, $E_{100}$=62 μeV/atom, respectively, which might be synthesized experimentally. The same projections of ID_6 towards $a$ and $c$ axes confirm their nearly same axial energies. The average distance of the first nearest neighbors of the N with its six Mn atoms is 1.97 Å. The average Mn atomic distance between the two neighboring layer is 2.36 Å. This layer structure is similar with that of non-magnetism structure (ID_2 in Table 3). Their difference is that the angle of the Mn$_6$N building blocks, it is about 90° in ID_6 but it is 83.7° in ID_2. Due to the not totally same Mn-N distance in Mn$_6$N building block, the MAE is thus formed in ID_6. Meanwhile, one structure ID_3 ($P2_12_12_1$, No.19) presenting similar energy with that of ID_2 is searched, whose $E_{001}$=36, $E_{100}$=119 μeV/atom, respectively.

A perfect easy-plane MAE structure ID_7 (Table 3) is searched, presenting large MAE values, $E_{001}$=270 μeV/atom, respectively, which shows diverse symmetries and has a $E_f$ of -0.1634 eV/atom. A similar structure is shown in ID_S21 in Table S9. Usually easy-axis MAE structures have more extensive applications than those of easy-plane ones, whereas easy-plane structures might also have potential applications in the future magnetic fields.

One structure ID_8 (Table 3) with $E_f$=-0.134 eV/atom and nearly perfect easy-axis MAE is predicted, whose $E_{001}$=60, $E_{100}$=58 μeV/atom, respectively. Its symmetry is dependent on the used tolerance, such as it is *P2/c* (No.13) at 0.001 Å, *Pmmn* (No.59) at 0.01 Å, *P4/nmm* (No.129) at 0.1 Å, respectively. A very similar case could also been seen in ID_9, which shows even lower energy than that of ID_8. ID_9 structure shows certain similarity with the structure ID_S20 (Table S9, $E_f$=-0.1399 eV/atom), whereas the MAE of latter is decreased, such as, $E_{001}$=5, $E_{100}$=38 μeV/atom, respectively, whereas both of which have same CN, namely, the N has six first-nearest Mn neighbors, and one of the Mn has five first-nearest N neighbors, whereas the other Mn has only one first-nearest N neighbors. Table S10 is the similar structure within energy ranges of about -0.1～0.11 eV/atom. A generally same structure (Table S11, ID_S22) with certain distortion is searched, presenting similar energy but smaller MAE compared with that of ID_8, in detail, whose $E_f$=-0.137 eV/atom and $E_{001}$=38, $E_{100}$=32 μeV/atom, respectively. Moreover, both structures have nearly same lattice parameters, indicating MAE is very sensitive on the atomic coordinates. Projections towards *a*,*c* orientations are nearly same, which causes the nearly same axial anisotropy energy.

The listed Table S12-14 is for reference purpose. A high-energy structure (Table S14, ID_S25) with $E_f$=-0.099 eV/atom presents nearly same MAE with respective values of $E_{010}$=72, $E_{100}$=74 μeV/atom, these MAE will transform to about 0, 117, 80 under small distortion (Table S14, ID_S26).

Another structure (Table 3, ID_10) shows nearly perfect easy-axis MAE with respective values of $E_{001}$=108, $E_{100}$=109 $\mu$eV/atom, in addition, its $E_f$=-0.112 eV/atom. Its N has six first-nearest Mn neighbors and its Mn has three first-nearest N neighbors. ID_11 is a simple lattice with medium MAE, differing with that of ID_2 which shows zero moment.

One structure (Table 3, ID_12) shows giant MAE, it is $E_{001}$=254, $E_{010}$=1376 $\mu$eV/atom, respectively, with symmetry $Pmc2$ (No.26) and $E_f$=-0.151 eV/atom, in which the magnetic moment are -0.6 and 1.36 $\mu_m$ for the two different kinds of Mn atoms, respectively. The values of the N are -0.002 $\mu_B$. Usually large moment induces large MAE, whereas it is not the case in this structure.

Two structures show large MAE, the maximum axial energy approaches to about 300 $\mu$eV/atom, its $E_f$ and symmetry (Table S12, ID_S23) are -0.154 eV/atom and $I41$ (No.80) at a tolerance of 0.001 Å, whose MAE are $E_{001}$=281, $E_{100}$=26 $\mu$eV/atom, respectively. Its N atom has six Mn neighbors, its Mn atoms has four or two N atoms, respectively. The other structure (Table S12, ID_S24) shows a nearly perfect easy-plane MAE with values of $E_{001}$=302, $E_{010}$=0.9 $\mu$eV/atom, respectively, which $E_f$=-0.106 eV/atom and symmetry is $Pmma$ (No.51) at a tolerance of 0.001 Å. Generally speaking, these two are the same structure.

A perfect easy-axis MAE structure (Table S16, ID_S27) is predicted, with $E_f$=-0.1 eV/atom, $E_{010}$=$E_{100}$=53.69 $\mu$eV/atom, respectively. Its symmetry is $P\bar{3}m1$ (No.164) and has two neighboring Mn layers, which is very similar with that of non-magnetism structure (Table 3, ID_2), whereas the present $E_{010}$=$E_{100}$=53 $\mu$eV/atom is caused by the lattice distortion as all of the $\alpha$, $\beta$, $\gamma$ have negligible variations, such as $α=90.0005°$, $β=89.9995°$, $γ=119.9995°$, respectively. It's structure is more similar with that of ID_11 in Table 3, whereas its energy is higher than those of ID_2 and ID_11 in Table 3.

Table 3 Experimental and searched low-energy Mn$_2$N structures information

| ID<br>$E_f$<br>Exp./Sea. | MAE<br>001<br>010<br>100 | Sym.<br>0.001 Å<br>0.01<br>0.1 | Unit cell<br>Mn<br>N | $a,b,c,$<br>$α,β,γ,$ |
|---|---|---|---|---|
| 1<br>-0.3323<br>Exp. | 59.9158<br>18.585<br>0 | $Pbcn$,60<br>$Pbcn$,60<br>$Pbcn$,60 | 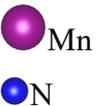 | 4.37186<br>4.80842<br>5.53472<br>90,90,90 |
| 2<br>-0.2451<br>Sea. | 0<br>0<br>0 | $P6_3/mmc$,194<br>$P6_3/mmc$,194<br>$P6_3/mmc$,194 | 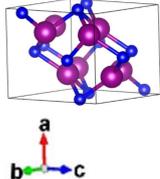 | 2.59805<br>2.59805<br>8.78183<br>90,90,120 |
| 3<br>-0.2457<br>Sea. | 36.2166<br>0<br>119.1833 | $P2_12_12_1$,19<br>$P2_12_12_1$,19<br>$P2_12_12_1$,19 | 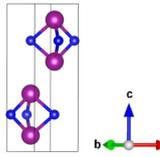 | 4.55135<br>5.81844<br>4.10135<br>90,90,90 |
| 4<br>-0.203<br>Sea. | 249.3958<br>0<br>250.0416 | $Fmmm$,69<br>$P4/mmm$,123<br>$P4/mmm$,123 | 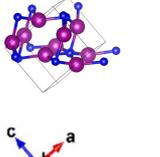 | 5.56460<br>7.67059<br>5.56905<br>90,90,90 |
| 5<br>-0.2295<br>Sea. | 10.2033<br>0<br>37.5916 | $P4_2/mnm$,136<br>$P4_2/mnm$,136<br>$P4_2/mnm$,136 | 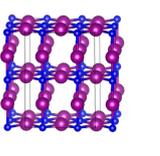 | 4.55210<br>4.55210<br>2.77969<br>90,90,90 |
| 6<br>-0.1736<br>Sea. | 62.4833<br>0<br>62.0333 | $Pccm$,49<br>$Pccm$,49<br>$Pccm$,49 | 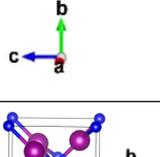 | 3.96094<br>7.05265<br>3.92921<br>90,90,90 |
| 7<br>-0.1634<br>Sea. | 270.0666<br>0<br>0 | Cmmm,65<br>Pmmm,47<br>P4/mmm,123 | 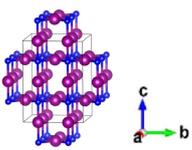 | 5.51706<br>5.54639<br>7.58013<br>90,90,90 |
| 8<br>-0.1345<br>Sea. | 60.2833<br>0<br>58.05 | P2/c,13<br>Pmmn,59<br>P4/nmm,129 | 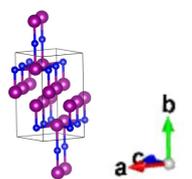 | 3.95580<br>7.01459<br>3.94784<br>90,<br>89.6052,<br>90 |
| 9<br>-0.1425<br>Sea. | 58.875<br>0<br>60.6333 | Pccm,49<br>P4/nmm,129 | 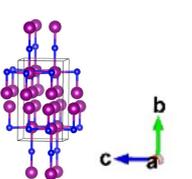 | 3.94971<br>7.02324<br>3.95589<br>90,90,90 |
| 10<br>-0.1125<br>Sea. | 108.7583<br>0<br>109.075 | Pna2$_1$,33<br>Pna2$_1$,33<br>Imma,74 | 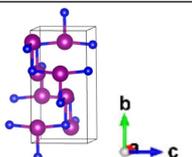 | 3.77101<br>7.44347<br>3.73550<br>90,90,90 |
| 11<br>-0.1979<br>available | 0<br>83.38<br>13.5766 | $P\bar{3}m1$,164<br>$P\bar{3}m1$,164<br>$P\bar{3}m1$,164 | 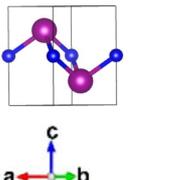 | 2.78588<br>2.78588<br>4.32834<br>90,90,120 |
| 12<br>-0.1516<br>Sea. | 254.2166<br>1,376.02<br>0 | Pmc2,26<br>Pmc2,26<br>Pmc2,26 | 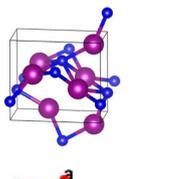 | 2.55864<br>4.34378<br>4.79211<br>90,90,90 |

**Mn$_5$N$_2$**

The energy difference between the searched lowest-energy structure (～-0.11 eV/atom) and the value at convex hull (～-0.24 eV/atom) is about 0.13 eV/atom, we thus only give a brief discussion for Mn$_5$N$_2$, as is shown in Table S17. The $E_f$ of the lowest-energy structure of Mn$_5$N$_2$ is about -0.11 eV/atom, which might be synthesized experimentally when its decomposed products are high-energy metastable Mn$_2$N and Mn$_3$N, as is shown in Figure 1. We didn't search Mn$_3$N owing to the absence of the experimental data, we thus ignored the detailed analysis to Mn$_5$N$_2$ despite its large MAE, such as one structure ID_S1 with $E_f$=-0.113 eV/atom shows very large MAE, it is $E_{010}$=252, $E_{100}$=215 $\mu$eV/atom, respectively, presenting a nearly perfect easy-axis MAE. The N is surrounded by six Mn, whereas Mn has three N neighbors with the average distance of about 2 Å.

**Mn$_4$N**

As is shown in Table 4 (ID_1), the symmetry and $E_f$ of the experimental Mn$_4$N are $Pm\bar{3}m$ (No.221) and -0.136 eV/atom, respectively, with MAE of $E_{001}$=97, $E_{010}$=12 $\mu$eV/atom, respectively. Several structures ID_S1-S5 show even lower energy than that of experimental one, whereas they present negligible magnetism, thus we ignore to discuss them, as is shown in Table S18 (Table S19-S22 are also the information for Mn$_4$N). In fact, ID_S1, ID_S3, ID_S4 are almost the same structure with different degree of distortion, ID_S2 and ID_S5 are also almost the same with that of experimental structure under some distortion. Despite that the chemical stoichiometry of Mn:N is 4:1, none of the searched low-energy structures shows very large MAE, the maximum value is about 200 $\mu$eV/atom, far smaller than those of MnN, indicating that the magnetism is not proportional to the magnetic atom content. Therefore, considering the more expensive price of Mn than that of N element, synthesization of cheap MnN or Mn$_2$N should be the first choice.

One Mn$_4$N structure ID_2 shows slightly higher energy ($E_f$=-0.129 eV/atom) than that of experimental one ($E_f$=-0.136 eV/atom), as is shown in Table 4, which shows a symmetry of *Fmmm* (No.69) and presents nearly perfect easy-axis MAE and medium values, $E_{001}$=126, $E_{010}$=121 $\mu$eV/atom, respectively, thus it is a very promising magnetic candidate material.

One Mn$_4$N structure ID_3 also presents nearly perfect easy-axis MAE with medium values and is only about 20 meV/atom higher ($E_f$=-0.1128 eV/atom) than that of experimental one, as is shown in Table 4, whose symmetry is *Imm2* (No.44) under a tolerance of 0.001 Å or *I4/mmm* (No. 139) under a tolerance of 0.1 Å, the correspondent MAE values are $E_{010}$=127 and $E_{100}$=133 $\mu$eV/atom. An iso-energy structure ID_S1-2 listed in Table S22 shows even larger MAE, with values of $E_{010}\approx$ 205, $E_{100}\approx$ 130 $\mu$eV/atom, respectively, behaving a symmetry of *I4/mmm* (No. 139) under all different kinds of tolerance 0.001, 0.01, 0.1 Å. Very small volume change ($a=b$=3.72534/3.72484, $c$=7.44036/7.43791 Å) doesn't change the lattice energy and MAE between ID_S1 and ID_S2. However, when the lattice parameters change to $a$=3.72588, $b$=3.72621, $c$=7.43791 Å, the $E_{010}$ reduced evidently, from 205 to 127 $\mu$eV/atom (shown in Table S22 and ID_S3), meaning that the variation range has exceeded the spin-orbital coupling interaction zone.

One Mn$_4$N structure ID_4 presents totally perfect easy-axis MAE and relatively large values, as is shown in Table 4, with values of $E_{001}$=$E_{100}$=169 $\mu$eV/atom, respectively, thus it is one of the ideal candidate materials and absolutely deserved to be synthesized experimentally. It is about 30 meV/atom higher than that of experimental data with $E_f$=-0.1 eV/atom. The similar structures and MAE induced by the slight distortion are shown in Table S19 ID_S6-7, in particular, ID_S7 has almost identical energy with that of experimental data and nearly perfect and large easy-axis MAE. Despite that ID_S8 shows slightly higher energy than that of ID_S7, it also shows nearly perfect and large easy-axis MAE, similar to that of ID_S7. In a word, many important metastable phases of Mn$_4$N could be synthesized and deserved to be carefully studied experimentally.

Table 4 Experimental and searched low-energy Mn$_4$N structures information,

| ID<br>$E_f$<br>Exp./Sea. | MAE<br>001<br>010<br>100 | Sym.<br>0.001<br>0.01<br>0.1 | Unit cell<br>●Mn<br>●N | $a,b,c,$<br>$α,β,γ,$ |
|---|---|---|---|---|
| 1<br>-0.1363<br>Exp. | 97.982<br>12.654<br>0 | $Pm\bar{3}m$,221<br>$Pm\bar{3}m$,221<br>$Pm\bar{3}m$,221 | 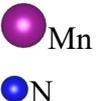 | 3.72357<br>3.72357<br>3.72357<br>90,90,90 |
| 2<br>-0.1292<br>Sea. | 126.325<br>121.475<br>0 | $Fmmm$,69<br>$Fmmm$,69<br>$Fmmm$,69 | 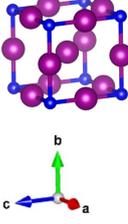 | 5.28113<br>10.38140<br>7.32419<br>90,90,90 |
| 3<br>-0.1128<br>Sea. | 0<br>127.408<br>133.934 | $Imm2$,44<br>$Imm2$,44<br>$I4/mmm$,139 | 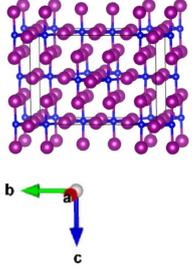 | 3.72588<br>3.72621<br>7.43791<br>90,90,90 |
| 4<br>-0.1004<br>Sea. | 169.19<br>0<br>169.19 | $I4/mmm$,139<br>$I4/mmm$,139<br>$I4/mmm$,139 | 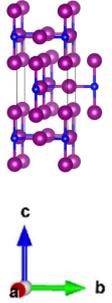 | 5.27033<br>7.40067<br>5.27023<br>90,90,90 |

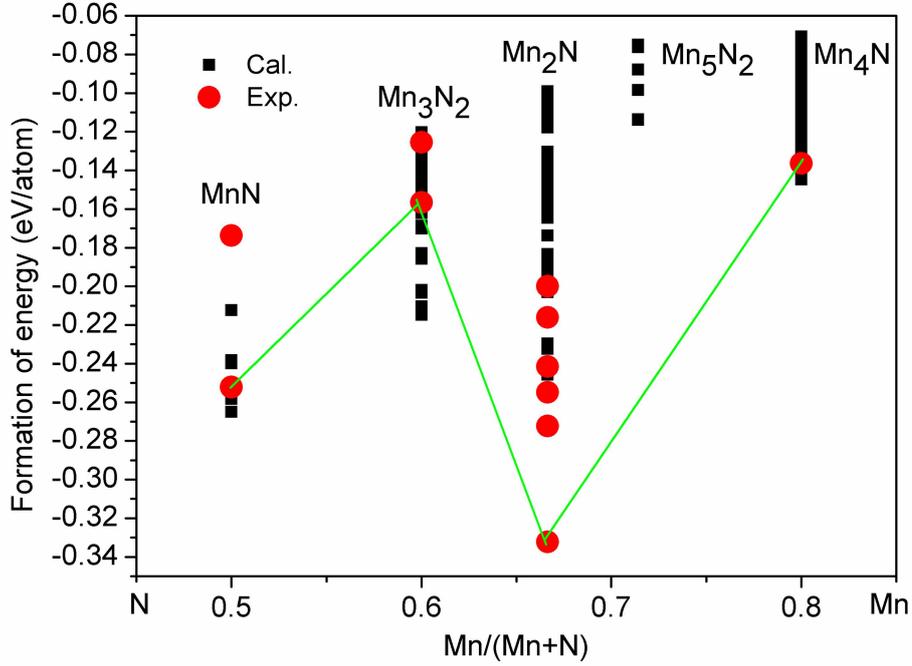

Figure 1. The convex hull of the manganese nitride, in which the alpha[4] Mn $I\bar{4}3m$ (No.217) and N$_2$ $Pa\bar{3}$ (205) is used to plot this figure, $P2_13$ (198) of N$_2$ has identical energy with that of $Pa\bar{3}$. Two convex hull for these compounds under different temperatures[14] and pressures[15] are plotted, respectively. Considering that alpha Mn $I\bar{4}3m$ (No.217) has 58 atoms in the unit cell, thus we ignore its MAE calculations, we also ignore the MAE calculation for the second lowest-energy structure as it is about 64 meV/atom higher than that of $I\bar{4}3m$.

**IV. Common features of these compounds with different stoichiometries**

There are 58 atoms in the ground state alpha Mn $I\bar{4}3m$ (No.217) and 20 atoms in $P4_132$ (No.213), the energy difference between above two element is about 0.06 meV/atom, which are the two lowest-energy structures. Our extensive search finds that the alloys of the binary manganese nitrides generally show simple unit cell, and the known and partial available experimental data are metastable states compared with the searched structures, originating probably from the complicated interaction of the magnetic Mn atoms during experimental synthesization. Therefore, the low-energy structures inherit their parent manganese structures during the doping of the nitrogen.

## V. Conclusions

Our extensive search found many perfect or nearly perfect easy-axis MAE structures, which are synthesizable experimentally and presents very useful applications in modern industry. It is very necessary for experimental researchers to synthesize them as present manganese nitrides are not very expensive in the market. The main motivation of this study is to find potential applications materials in manganese nitrides in the field of spin-electronics. The magnetic properties thus could be deeply studied once the structure is synthesized experimentally. The general dependence of the MAE on the distorted structure is carefully presented. Many large MAE structures are searched in MnN despite its less Mn conent, deviating the basic knowledge that the more magnetic atoms have, the large MAE will be. A nearly perfect easy-axis giant MAE structure in MnN is searched and also is synthesizable experimentally, presenting unprecedentedly giant MAE in this new energy materials field. Several large nearly perfect easy-axis MAE structures in $Mn_3N_2$ metastable states are searched, which are also synthesizable experimentally, presenting important potential applications in the magnetic fields. We also searched several almost totally perfect easy-axis MAE structures in the $Mn_2N$ metastable states, these phases have great promising to be synthesized by the non-equilibrium technology. We searched one nearly perfect easy-axis MAE structure in $Mn_5N_2$ metastable states that is possible to be synthesized when considering its decomposed products are metastable $Mn_2N$ and $Mn_3N$. Finally, three perfect or nearly perfect easy-axis MAE structures in the $Mn_4N$ metastable states are searched. In a word, these about ten important structures have extensive application promising in the spin-electronics fields.